
\input phyzzx.tex
\input phyzzx.fonts

\def\bp{{\bf p}}
\def\bq{{\bf q}}
\parskip=10pt
\frontpagetrue
\rightline{UCTP-103/1992}
\vskip1in
\centerline{\seventeenrm  A New Field Theoretic Approach}
\centerline{\seventeenrm to Criticality }

\vskip 0.3in

\centerline{P. SURANYI\footnote\dagger{E-mail:
suranyi@ucbeh.san.uc.edu}\footnote\*{PACS numbers 11.10Ef and 64.60.Fr}}

\centerline{\it Department of Physics, University of Cincinnati, }
\centerline{\it Cincinnati, Ohio 45221-0011, USA }
\vfill

\centerline{\bf Abstract}
{\narrower\narrower\smallskip\noindent A reorganized perturbation expansion
with a propagator of soft infrared behavior is used to study the critical
behavior of the mass gap. The condition of relativistic covariance fixes the
form of the soft propagator. Finite approximants to the correlation critical
exponent can be obtained in every order of the modified, soft perturbation
expansion. Alternatively, a convergent series of exponents in large orders of
the soft perturbation expansion is provided by the renormalization group in all
spatial dimensions, $1\leq D\leq3$. The result of the $\epsilon$-expansion is
recovered in the $D\rightarrow 3 $ limit.
\smallskip }
\eject
\chapter{Introduction}
The critical behavior of various statistical systems near
continuous phase transitions has been the subject of intense
investigations for several decades. Continuum quantum field
theory turned out to be the most appropriate tool for these
investigations. It connects critical phenomena with the
fundamental features of renormalization and renormalization
group. Nevertheless, the best numerical values for critical
parameters are obtained more often then not from lattice
methods: high and low temperature expansions, Monte-Carlo
investigations, etc. The reason for this dichotomy
 is obvious. Numerical methods on lattices are convergent
procedures, which, as a function of parameters (or computer
time), diverge only exactly at the critical point. In
contrast to that, the critical behavior in continuum field
theories is investigated through the $\epsilon$-expansion of
Wilson and Fisher\Ref\fisher{K.G.Wilson and M.E.Fisher \journal Phys. Rev. Lett
& 28 (72) 240;}, which define only an
asymptotic series for all values of the expansion parameter,
$\epsilon$.\Ref\brezin{E. Brezin, J-C. Le Guillou, and J. Zinn-Justin \journal
Phys. Rev. & D15 (77) 1544;}

In the current paper, we intend to introduce a new method for
the investigation of critical behavior in quantum field
theories. The method is based on a reorganization of the
perturbation expansion series, very similar to mass
renormalization, with the intent of extracting key
nonperturbative features from field theories. The purpose of
the modification of the propagator is to soften the infrared
behavior of the propagator, thus providing a finite
approximant to the correlation critical exponent in every
order of the perturbation expansion. The condition of
relativistic covariance fixes the form of the softened
propagator up to a scale transformation. The form of the
propagator is simply related to the correlation
exponent, $\nu$. The series of approximants for $\nu$,
obtained from successive orders of the soft perturbation
expansion, is convergent, or at least has a convergent subseries. In this
respect, our method is
unique.

The method will be demonstrated on the example of the
$\phi^4$. In the next section, the
modified perturbation expansion will be motivated
and defined. Sec. 3. will prove the existence of a soft
propagator consistent with the requirement of correct
relativistic form of the energy of single particle states.
The renormalization group approach will be applied to the theory defined by
the modified perturbation expansion in Sec.
4. We will prove that an alternative series of exponents can be obtained in
every spatial dimension, $1\leq D\leq3$. Near $D= 3$ the leading order
approximation reproduces the one loop result of $\epsilon $-expansion. In the
final section, our results will
be summarized.
\chapter{Modified Perturbation Expansion}
The stationary state Schr\"odinger equation for $\phi^4$
field theory in $D$ spatial dimensions is given by
$$H\ket{\psi}={1\over2}\int d^Dx
\left(\Pi^2(x)+m^2\phi(x)+\partial_\mu\phi(x)\partial_\mu
\phi(x)+{g\over12}\phi^4(x)\right)
\ket{\psi}=E\ket{\psi},\eqn\one$$
The investigation of a field theory in Schr\"odinger
representation is easiest using the representation
$$\Pi(x)={1\over i}{\delta\over\delta\phi(x)}\eqn\three$$
for the canonical momentum operator.
Then one can choose $\ket{\psi}$ to be a functional of the
classical field $\phi$.\Ref\jackiw{R. Jackiw: ``Functional
Representation for Quantized Fields'', in {\it Proceedings of
the 1st Asia Pacific Workshop on High Energy Physics,
Conformal Field Theory, Anomalies and Superstrings}, ed. B.E.
Baaquie et al., World Scientific, Singapore, 1988;} This
corresponds to the coordinate representation of quantum
mechanics.

We will investigate the equations for the ground and first
excited (single particle) states simultaneously. Our aim is
to extract information concerning the critical behavior of
the renormalized mass. We will denote the energy eigenvalues
of the ground state and the first excited state by $E^{(0)}$
and $E^{(1)}=E(\bq)+E^{(0)}$, respectively. The first excited
state of the system is a single particle state. The excitation
energy of a single particle state must have the correct
relativistic form
$$E^2(\bq)=\bq^2+m^2,\eqn\eight$$
where the notation $m$ is used for the physical mass.

As usual, we rearrange the Hamiltonian by subtracting a mass
renormalization term,
$$\delta H={\delta m^2\over2}\int d^Dx \phi^2(x),\eqn\seven$$
from the kinetic part and adding it to the interaction part.
Then, with an appropriately chosen $\delta m^2=m_0^2-m^2$, the
perturbation expansion becomes finite for $D<3$.  We define
the subtraction scheme more precisely below.

The Hamiltonian perturbation expansion (or old fashioned
perturbation expansion) is completely equivalent to the
covariant perturbation expansion. After collecting terms in
every order of the perturbation expansion, one can introduce
the extra $p_0$ momentum components and get an expression for
the squared energy function $E^2(\bp)$ as
$$E^2(\bp)=\bp^2+m^2+\Sigma(p_0,\bp)\bigg|_{p_0^2=
\bp^2+m^2},\eqn\extra$$
where $\Sigma(p_0,\bp)$ is the sum of self-energy
corrections. The self-energy correction is covariant,
thus, it depends on $ p^2=p_0^2+\bp^2$ only, but for future
reference we indicate its dependence on the energy component
independently. Furthermore, if $m$ is the renormalized mass,
the self-energy correction must be subtracted as
$\Sigma(m,0)=0$. Then, of course, every term of the covariant perturbation
expansion of $\Sigma$, being momentum independent, vanishes
for all $\bp$. The reader may ask at this point, why do we go
through this elaborate procedure to define an identically
zero quantity? The answer is that we will use a similar
procedure later, in which the vanishing of $\Sigma$ will not
be trivial at all, and at that point we will draw on the
analogy to the trivial excercise above.

The information concerning the critical behavior extracted from the
perturbation expansion is of
a divergent nature. The root of this problem is
shown clearly by the $m$ dependence of the terms of the
perturbation expansion of $\delta m^2$. Dimensional
considerations show that in $n$th order of the perturbation
expansion, after the removal of a term diverging when the
cutoff, $\Lambda\rightarrow\infty$, but independent of $m$,
one is left with a term depending on the renormalized mass as
$$\delta m^2=c_ng^n\mu^{n(3-D)}m^{2-(3-D)n},
\eqn\eight$$
where $c_n$ is a dimensionless constant. For every $1\leq D<3$ and
for sufficiently large $n$, the power of $m$ becomes negative.
On the other hand, the dependence of the `exact' $\delta m^2$
on $m$ is related simply to the critical behavior
$$\delta m^2=m_c^2(\Lambda)+cm^{1\over
\nu}+O(m^\lambda),\eqn\extraoneandahalf$$
where $\lambda>1/\nu$ and $\nu$ is the correlation critical
exponent and $m_c^2(\Lambda)$ is the critical value of the squared bare mass.
Thus, the $m\rightarrow 0$ behavior of \eight\ has
nothing to do with the real critical behavior.

Our thesis is that one does not obtain the correct critical
behavior because some crucial non-perturbative features of
the theory are overlooked by a straightforward perturbative
expansion. Our aim is to incorporate some non-perturbative
features into the expansion such that the critical behavior
of the renormalized mass can be extracted. The procedure
we intend to employ is very similar to the additive
renormalization of mass.

There are examples of dramatic improvement of the convergence
of perturbation series by reorganizing the perturbation
series. Halliday and Suranyi were able to obtain convergent
perturbation expansions for the anharmonic oscillator in
quantum mechanics by an appropriate reorgnization of the
series.\Ref\halliday{I.G. Halliday and P. Suranyi \journal
Phys. Lett. & 85B (79) 2134;}\Ref\hallidaytwo{I.G. Halliday and
P. Suranyi \journal Phys. Rev. & D21 (80) 1529;} Following the
spirit of these works, we will alter the propagator by adding
a term
$$\delta H={1\over2}\int d^Dp
\Delta(\bp)\phi^\dagger(\bp)\phi(\bp)\eqn\ten$$
to the kinetic part of the Hamiltonian and subtracting it
from the potential part. By this rearrangement, we are not
able to make the perturbation series convergent for
non-universal quantities like the coefficient of the term
$m^{1/\nu}$, but we are able to extract convergent
approximations for the exponent $\nu$.

At this point, the form of the corrected propagator is
undetermined. Using an arbitrary propagator, however, changes
the perturbation series in an uncontrolled manner, thus, the series
expansion for $E(\bp)$ will not have the correct
relativistic form \eight. We will show in the next section
that requiring the correct relativistic form for $E^2(\bp)$ in
every order of the perturbation expansion fixes the form of
$\Delta(\bp)$, up to an irrelevant overall multiplier.

It will be convenient to use the $D+1$ dimensional,
`relativistic' form of the perturbation expansion.
Then the propagator will take the form
$$D(\bp,p_0)= {1\over
p_0^2-\omega^2(\bp)+i\epsilon},\eqn\egy$$
where
$$\omega^2(\bp)=\bp^2+m^2+\Delta(\bp),\eqn\egyesfel$$
while the interaction term will be
 $$\eqalign{L_{\rm int}&={1\over2}\int
d^{D+1}x\left[{g\over12}\phi^4(x)+\delta
m^2\phi^2(x)\right]-{1\over2}\int d^{D+1}p\Delta(\bp)
 |\phi(p)|^2\cr&=L_{\rm int}+L_{\delta m^2}+L_\Delta,}\eqn\ketto$$
where for convenience we have given $L_\Delta$ in momentum
representation.

It is easy to see, starting from the $D$ dimensional, Hamiltonian form of the
theory, that in the language of the $D+1$ dimensional `relativistic'
theory, the quantity we must calculate is
$$E^2(\bq)=\omega^2(\bq)+\Sigma(q_0,\bq)
\bigg|_{q_0^2=
\omega^2(\bq)}.\eqn\extratwo$$
In contrast to the similar expression obtained in the standard perturbation
expansion, \extra, one does not
trivially obtain the correct relativistic form for the single
particle energy. Requiring the correct relativistic form \eight\ implies the
nontrivial integral
equation
$$\Delta(\bq)+\Sigma(q_0,\bq)
\bigg|_{q_0^2=
\bq^2+m^2+\Delta(\bq)}=0\eqn\extrathree$$
for $\Delta(\bq)$.\footnote1{In a previous work, we obtained results for
critical exponents, identical to the ones obtained in this paper, by a
different interpretation of \extratwo.\Ref\suranyi{P. Suranyi, preprint UCTP
106/91, to be published;} We will comment on that interpretation in Sec. 5.}

There is a trivial solution of \extrathree, $\Delta(\bq)\equiv0$, reducing
\extrathree\ to \extra. We will prove in the next section that a nontrivial
solution for $\Delta(\bq)$ exists as well such that in every order of
the modified perturbaton expansion \extrathree\ is satisfied and the correct
dispersion
relation for the energy is maintained.
\chapter{Existence of the Modified Expansion}
The perturbative terms become bounded functions of the physical mass, $m$,
unlike \eight, only if the infrared behavior of the propagator is softened.
Thus $\Delta(\bp)$ should have a leading
order infrared behavior
$$\Delta(\bp)\simeq |\bp|^{2\alpha}\xi^{2-2\alpha},\eqn\harom$$
where $0<\alpha<1$ and $\xi$ is a scale parameter of the
dimension of a mass.
Then the $\bp$ dependence of the propagator is dominated by
$\Delta(\bp)$ in the interval $I$,
$$I:\xi>>|\bp|>>m.\eqn\negy$$
The euclidean propagator of the `relativistic' perturbation
theory has the form
$$D(p_0,\bp)={1\over p_0^2 +\bp^2+\xi^{2-
2\alpha}|\bp|^{2\alpha}+m^2+...}.\eqn\negy$$

In interval $I$, the propagator can be approximated by
$$D(p_0,\bp)={1\over p_0^2+|\bp|^{2\alpha}}, \eqn\ot$$
where we absorbed the scale factor into momentum $\bp$ (in
other words, we set $\xi=1$).

For the time being, we will omit term $L_\Delta$ of \ketto\ in
the perturbation expansion. We will prove later that its
inclusion would not change our results for the critical
exponents.

It will be sufficient to study \extrathree\ in
interval $I$ to fix the form of $\Delta(\bp)$. We can argue
that if we wish to extract the leading behavior of
the terms of our expansion while the external momentum is in
interval $I$, then it is self consistent to use  propagator
\ot,
instead of the complete form \negy. That is certainly so if
the
loop integrals are dominated by contributions in which all
internal momenta are in interval $I$ as well.

An $n$ loop
contribution generated by such propagators has the behavior
$|\bq|^{a_n}$, where $a_n={2\alpha+n(D-3\alpha)}$. This can
be
seen if we realize that in a $\phi^4$ theory there are $2n-1$
propagators, each contributing $|\bp|^{-2\alpha}$, while the
momentum integrations over $\bp$ and $p_0$ contribute
$|\bp|^{Dn}$ and $|\bp|^{n\alpha}$,
respectively\footnote1{Note
that $p_0=O(|\bp|^\alpha)$ in loop momenta.}. Using similar
arguments, it is easy to
see that $n$ loop vertex corrections have the
overall momentum dependence $|\bq|^{b_n}$, where $b_n=n(D-3\alpha)$. Now a
necessary condition for the finiteness of self
energy corrections is the finiteness of
vertex corections. Those are finite only if
$$\alpha>\alpha_0={D\over3}.\eqn\nyolc$$
We will restrict ourselves to such values of $\alpha$ in the
future.

Suppose now that in a certain loop integral the loop momentum
is
in the region $|\bp|\geq\mu$, but the momenta external to the
loop
are in $I$. Then the loop is effectively shrunk to a point
vertex
with at least four external legs (self energy
insertions will be discussed below). If the loop contains
$k$ propagators ($k\geq2$),
then the overall behavior of the diagram has been {\it
increased}
by a power of $2k\alpha-\alpha-D>0$, in view of the
constraint
\nyolc\ on $\alpha$. Then we do not obtain a leading
contribution, since $|\bq|<<\xi=1$.

If the external
momentum of a self energy insertion is in $I$, but internal
momenta are not, then the correction is even smaller. This
can be seen by observing that self energy corrections are
subtracted at $p_0^2=\bp^2=0$ ($m^2$ can be neglected in $I$) and the largest
contribution
must be proportional to $p_0^2$. Then
observing that $p_0^2=O(|\bp|^{2\alpha})>>|\bp|^{a_k}$, for
all $k\geq1$, we can see that again we obtain a non-leading
contribution. Thus we have proved that if the external momentum is in interval
$I$, then propagator (3.4) should be used.

In $N$th order of the modified expansion, the infrared behavior of the
diagrams is controlled by power $a_n=2\alpha+n(D-3\alpha)$.
If
$a_n>0$, diagrams are infrared convergent, even at vanishing external momentum.
We can easily see
that the subtraction at $q_0=\bq=0$ makes the contribution
finite
as long as $b_n=n(D-3\alpha)<0$, where ultraviolet
divergences
will appear. Then the subtracted $n$ loop integral is finite in interval
$J_n$
$$J_n:~~{D\over 3-{2\over n}}>\alpha>{D\over3}.\eqn\kilenc$$
In the subtraction term, we set $q_0=0$ and $\bq=0.$ The
subtraction results in differences having large momentum
behavior
$$|\bp-\bq|^{2\alpha}-|\bp|^{2\alpha}=|\bp|^{2\alpha-
2}\left[\alpha|\bq|^2 -2\alpha\bp\cdot\bq+2\alpha
(\alpha-1){(\bq\cdot\bp)^2\over \bp^2}+...\right],\eqn\tiz$$
where $\bp$ is a loop momentum. Since the leading correction term,
$-2\alpha\bp\cdot\bq$,
cancels after angular integrations over $\bp$, the ultraviolet power behavior
is
reduced by two. The subtraction in
an
expression of the form $(p_0-q_0)^2-p_0^2=q_0^2-2p_0q_0$
reduces
the power behavior by $p_0^2\sim|\bp|^{2\alpha}$
only\footnote1{Note that $|\bp|^{2\alpha}>>\bp^2$, if
$|\bp|\in I$.}. In other words, the ultraviolet behavior of the subtracted
self-energy correction term is
controlled by the power $a_n-2\alpha=b_n$. This is, of couurse, the same power
as the one controlling the ultraviolet behavior of the vertex part. Our
assertion
concerning the convergence
of the term of the expansion in interval  $J_n$, \kilenc, has been
proved.
 \section{Existence of a nontrivial solution}
Next we show that a value for $\alpha\in J_n$ can be found such
that the leading term in \extrathree\ vanishes. Non-leading
order terms that might appear in \extrathree\ and non-leading terms of
$\Delta(\bq)$ will be considered later.
Suppose we work in $n$ loop order of our approximation
scheme. Then if $|\bq|$ is in interval $I$, the largest order
($n$ loop) term dominates. This is so, because the series of
exponents $a_k$ satisfies $a_k>a_{k+1}$.
Since $|\bq|^{2\alpha}<<|\bq|^{a_n}$, solution \harom\ of  \extrathree\ is
possible only
if the
coefficient of the term $|\bq|^{a_n}$ vanishes.

First, in a given order of the loop expansion all euclidean diagrams are real
and have the same sign. The subtraction (mass renormalization) term
of
every diagram becomes infrared singular at $a_n$=0. Since the
subtraction term is always negative, the $n$ loop term tends to
$-\infty$ if $\alpha\rightarrow D/(3-2/n).$ On the other
hand, it
is also easy to see that the $n$-loop contribution tends to
$+\infty$ at $\alpha\rightarrow D/3$, the point of
ultraviolet
divergence. The contributions are analytic functions of the
variable $q_0^2$. In other words, we can write a dispersion
relation for the subtracted contribution
$$\int^\infty dz \left[{f(z,|\bq|)\over q_0^2+z}-{f(z,0)\over
z}\right],\eqn\tizenketto$$
where the positive definite discontinuity has an asymptotic
behavior $f(z,\bq)\sim z^{a_n/2\alpha}$. Note that
the constraint $0<a_n<2\alpha$ insures the convergence of
\tizenketto.

Dimensional considerations show that $|\bq|$
can be scaled out from \tizenketto\ to give
$$|\bq|^{a_N}\int^\infty dz
\left[{f(z,|\hat1|)\over
q_0^2/|\bq|^{2\alpha}+z}-{f(z,0)\over
z}\right].\eqn\tizenharom$$
Furthermore, according to our previous discussions, the
leading asymptotic behavior of the $n$ loop term comes from
the
term $q_0^2$ in the denominator of the integrand. Thus, at
large values of $z$, we can also set the
momentum $\hat 1$ equal to zero in the argument
of the spectral density. Finally, according to our
prescription,
we have to set $q^2_0=-\bq^2-m^2-\Delta(\bq)\simeq-|\bq|^{2\alpha}$, for the
Euclidean
$q_0$.
Then we obtain the following expression for the large $z$ contribution,
responsible for the ultraviolet divergence
$$|\bq|^{a_n}\int^\infty dz f(z,0){1\over z(z-
1)}.\eqn\tizennegy$$
\tizennegy\ shows that the integral indeed tends to $+\infty$ when
$\alpha\rightarrow D/3$. We can conclude that the $n$
loop
contribution to the subtracted self energy diagram has a zero at some
$\alpha_n\in J_n$.
At $\alpha=\alpha_n$, the coefficient of the leading
power,
$|\bq|^{a_n}$, where $a_n=2\alpha_n+n(D-3\alpha_n)$,
vanishes.

We turn now to the discussion of non-leading order terms in
$\Delta(\bq)$. Though by setting
$\alpha=\alpha_n$, the $n$-loop contribution vanishes, the
correct dispersion relation for $E^2(\bq)$ requires that the
coefficients of the
powers $|\bq|^{a^k_n}$ ($k$-loop term), where
$\alpha_n^k=2\alpha_n
+k(D-3\alpha_n)$, $k<n $ should also vanish. For
$k>0$, these powers are also smaller than $2\alpha$ and give
contributions much larger then $|\bq|^{2\alpha}$ in the region
$|\bq|<<1$, so they should also be cancelled in \extrathree. It is easy to see,
however, that a non-leading term of
$\Delta(\bp)$ in the propagators of the loop diagram can cancel these
contributions as well as the term $\Delta(\bq)$ appearing in \extrathree. Thus
$\Delta(\bp)$ should have the form
$$\Delta(\bp)=
\sum_{l=0}g^lc_l|\bp|^{2\alpha+l(3\alpha-D)}.\eqn\tizenot$$
Then an expansion around the leading contribution
$c_0|\bp|^{2\alpha}$ leads to correction terms to the $n$
loop
contribution, which have exactly the same behavior as the
$n-1$,
$n-2$, etc. loop terms. An appropriate choice of the
coefficients $c_l$ in \tizenot\ cancels these contributions.
It is
amusing to notice that in large orders of the perturbation
expansion the coefficients $c_i$ decrease very fast with $n$,
e.g.
$c_1\sim c_0^3/n^2$.

Finally, let us return to the neglected term, $L_\Delta$, of
the perturbation expansion. It is easy to show that the
contributions of $L_\Delta$ are similar to those of $k$-loop terms, $k<n$, of
the perturbation
expansion. Indeed, if we work in $n$ loop order, an insertion
$\Delta(\bp)$ on a propagator would substitute a four
point vertex, lowering the power of $g$ by one, and would
substitute one of the propagators $D(q_0,\bp)$ by $D(q_0,\bp)
\Delta(\bp)D(q_0,\bp)=O(D(p_0,\bp))$. In other words, we
obtain a contribution with the same power of $g$ and the same
low momentum behavior as the $n-1$ loop term. We know,
however, that the $n$ loop term provides the dominant power of
the expansion, thus the equation determining $\alpha$ is not
affected by the term $L_\Delta$ of $L'$, defined in \ketto. These non-leading
terms can also be cancelled in \extrathree\ successively, by non-leading terms
of the asymptotic expansion of $\Delta(\bp)$.

\section{Critical exponent}

The correlation critical exponent $\nu$ relates the behavior of the
unrenormalized
mass and the renormalized mass
$$m_0^2=m^2+\delta m^2=m^2+f(\Lambda)+cm^{1\over\nu },\eqn\tizenhat$$
where $f(\Lambda ) $ is the cutoff dependent critical point
for
the square of the bare mass and $c$ is a constant. The exponent $\alpha_n$ has
a direct connection with the
$n$th
order approximant of  $\nu$. One is able to extract $\nu$ if one investigates
the behavior
of
the terms of the modified expansion for $\delta m^2$ in the transition region,
$|\bq|\simeq
m$.  In
that region, the mass term cannot be neglected and we have to
use the form $p_0^2+|\bp|^{2\alpha}+m^2$ for the propagator.

Due to the
infrared convergence of the integrals in interval \kilenc\
we
can separate terms of the form $c_kg^k\Lambda^{\alpha_n^k}$
from the mass renormalization constant, such that the
remainder is a finite function $m$. Then we can scale out
the mass, $m$, from the terms of the perturbation expansion
with the simultaneous
substitutions
$\bp\rightarrow m^{1\over\alpha}\bp$ and $p_0\rightarrow
mp_0$.
Using \tizenhat, we obtain the following expression for
$\nu_n$, the estimate for the critical exponent  $\nu$ in
$n$-loop approximation,
$${1\over\nu_n}=2-n(3-{D\over\alpha_n}).\eqn\tizennyolc$$

Equation \tizennyolc\ shows that the dominant contribution (smallest
power
of $m$) comes from the leading order, $n$-loop term, since  for $k<n$ we have
$2-k(3-D/\alpha_n)>2-n(3-D/\alpha_n)$, provided $\alpha\in J_n$ of
\kilenc. Furthermore, it also follows from \kilenc\ that the
value of $\nu$ is between $1\over2$ and $\infty$, as it
indeed
should be.

The natural method for finding approximants to
$\nu$ would be to find values for $\alpha_n$ at as high $n$
as
possible. Indeed, in  a previous paper we found the critical
exponent for $D=1$ and $D=2$ in the two-loop
approximation, $\nu^{(1)}=1.013$, $\nu^{(2)}=0.713$, in reasonable agreement
with known
values.\refmark\suranyi

There is a puzzling question concerning our approximation scheme.
Since $\alpha_n$ is obtained from $n$-loop diagrams only,
completely different from the $(n-1)$-loop diagrams used for obtaining
$\alpha_{n-1}$,
it is hard to see
the relation between subsequent
approximants to $\nu$, $\nu_n$ and $\nu_{n-1}$. It could happen that
$\nu_n$, as a function of $n$, oscillates
between ${1\over2}$ and $\infty$, instead of being convergent. A bounded
infinite series always has at least one convergent subseries, but it may have
several that converge to different limits.
As is shown by \tizennyolc, a smooth limit to the correct value of $\nu$
is possible only if we can prove that in
large orders of
the loop expansion the equation determining $\alpha$ becomes
a  function of $\rho=n(3-D/\alpha)$ only, thus making the
existence of a
nontrivial limit $\rho\rightarrow\rho_c$ for
$n\rightarrow\infty$
possible. Then, of course, $\nu={1\over2-\rho_c}$.

The simplest method for the investigation of large orders
of the perturbation
expansion is the renormalization group.
Observe that the modified loop expansion is equivalent to a
perturbation expansion in an euclidean field theory with
action\footnote1{For the sake of simplicity, we mix coordinate space and
momentum space
representations for $\phi$. The action
could be written in either of these representations without
any difficulty.}
$$\eqalign{S(\phi)&={1\over2}\int dp_0
d^Dp[p_0^2+|\bp|^{2\alpha}]
\phi(p)\phi(-p)\cr&+{1\over4!}\int dx_0d^Dx\phi(x)^4+{\delta m^2\over2}\int
dp_0
d^Dp
\phi(p)\phi(-p),}\eqn\husz$$

Similar field theories have been investigated in the past.
The
closest example is that of Fisher and Grinstein\Ref\fishertwo{M.
P. Fisher and G. Grinstein \journal Phys. Rev. Lett.& 60 (88) 208;} who
found that certain superconducting systems may be
described by Hamiltonians similar to \husz. They, however use
a
`double $\epsilon$-expansion' in the variables $\epsilon=3-D$
and
$\epsilon'=1-\alpha$. Our intention is to investigate field
theory \husz, at fixed $D$ near the critical point
$\alpha=D/3$.
In other words, we wish to perform an expansion in
$\delta=3\alpha-D$.

We will show in the next section that the Callan-Symanzik
equation may be used to find the coefficients of terms,
singular in
$\delta$, in large order contributions to Green's function.
This is quite analogous to the calculation of leading
logarithms in perturbation theory. We
will also find that up to a multiplier, independent of the  momentum,
the
large order contribution to the self energy correction
becomes a
function of the variable $\rho=3n\delta/D$. Higher and higher
orders of the loop expansion provide a power series of higher
and
higher order in variable $\rho$. Finding the zero of the self
energy  correction as a function of $\rho$ provides an
alternative method for the determination of $\nu$. One
outstanding feature of the power series expansion in $\rho$
is that in contrast to the $\epsilon$-expansion, it has a
finite radius of convergence.

\chapter{Renormalization Group}
\section{Callan-Symanzik equations}
The renormalization of the field theory, defined by action
\husz\ requires wave function and coupling constant
renormalizations, besides the additive mass renormalization we discussed
earlier. The renormalized and bare amputed
Green's functions have the following relation with each other
$$\Gamma^{(4)}(p_1,p_2,p_3,p_4,g,\mu)=Z^{-
2}(g,\delta)\Gamma^{(4)}_0
(p_1,p_2,p_3,p_4,g_0,\delta),\eqn\fiftyone$$ and
$$\Gamma^{(2)}(p,g,\mu)=Z^{-1}(g,\delta)\Gamma^{(2)}_0(p,g_0,
\delta),\eqn\fiftythree$$
where $\delta=3\alpha-D$ and $\mu$ is the renormalization
scale, defined through the relation of the renormalized and the
unrenormalized charges
$$g_0=g\mu^{\delta}Z_g(g,\delta).\eqn\fiftytwo$$
$p_i$ denote both the spatial, $D$ dimensional
part of the momenta, and their timelike part as well.
$\Gamma^{(2)}$ is the inverse propagator, the main
subject of our investigation.

The derivatives of the left hand sides of \fiftyone\ and
\fiftytwo\
with respect to $1/\delta$ should vanish
as $\delta^2$, when $\delta\rightarrow0$, because their first
nonvanishing $\delta$-
dependent terms are proportional to $\delta$. Thus the
unrenormalized Green's functions satisfy the following
equations
$$\left[{\partial\over\partial1/\delta}+\beta(g,\delta)
{\partial\over \partial g}-
2\gamma(g,\delta)\right]\Gamma^{(4)}_0(p_1,...
p_4,g,
\delta)=O(\delta^2),\eqn\fiftyfour$$
and
$$\left[{\partial\over\partial1/\delta}+\beta(g,\delta)
{\partial\over\partial g}-\gamma(g,\delta)\right]
\Gamma^{(2)}_0(p,g,\delta)=O(\delta^2), \eqn\fiftyfive$$
where we dropped the subscript 0 of the coupling constant.  Strictly speaking,
it is not exactly $\Gamma^{(2)}$ that satisfies \fiftyfive, as it will soon
become obvious.
 The
relation between the standard cutoff dependent form of the
Callan-Symanzik equations and the above forms is given by
the two possible limits
$$\ln(\Lambda/\mu)\leftarrow{1-(\Lambda/\mu)^{-\delta}
\over\delta}\rightarrow {1\over\delta}.
\eqn\fiftyeight$$
While the cutoff scheme is obtained if one takes the limit
$\delta\rightarrow0$ first, the dimensional regularization
scheme is obtained if the limit $\Lambda\rightarrow \infty$
is taken first.
The beta and gamma functions can be determined from
\fiftyfour\ and \fiftyfive, taken at
specific momentum values. In the $\delta\rightarrow0$ limit
they coincide with the standard definitions.

The renormalization group functions have the following
perturbation expansion
$$\beta(g)= \beta_2g^2+\beta_3g^3+...\eqn\fiftynine$$
and
$$\gamma(g)=\gamma_2g^2+\gamma_3g^3+...\eqn\sixty$$

\section{Calculation of self energy corrections in large orders of the soft
perturbation expansion}

The two point amplitude can be written in the form
$$\Gamma^{(2)}_0(p_0,\bp,g,\delta)=p^2_0+|\bp|^{2\alpha}+p^2_0
\Xi(p_0 ,\bp,g,\delta),\eqn\sixtyone$$
where we explicitly indicated the separate dependence on
$p_0$ and $\bp$. The important property of $\Gamma^{(2)}$ is
that the coefficient of $|\bp|^{2\alpha}$ does not get
renormalized.
When the subtraction at zero momentum is performed in the loop
integrals, the leading terms in the difference are of the
form
$(p_0-q_0)^2+(p_0+q_0)^2-2q^2_0=p_0^2$ and $(\bp-\bq)^{2\alpha}+
(\bp+\bq)^{2\alpha}-2\bq^{2\alpha}\sim \bp^2<<p^2_0$, if $\bp<<\xi$.
That of course shows that indeed, the coefficient of
$\bp^{2\alpha}$
does not get renormalized, and we should drop the second term
from the right hand side of \sixtyone, when we substitute into
\fiftythree\ or \fiftyfive.

We expand the function $\Xi $ in $g$ as follows
$$\Xi(p_0,\bp, g,\delta)=\sum_n {  (-\beta_2g)^n\over
\delta^{n-1}}Q_n(\delta,p), \eqn\bone$$
where the function $Q$ has the series expansion
$$Q_n(\delta,p)=\sum_{m=0}^{n-1}\delta^m q^m_n(p).
\eqn\btwo$$
The leading coefficient $q^0_n(p)=1$, by the normalization we
chose for $Q_n$, while $q^{n-1}_n(p) $ is the finite part of the $n$ loop
amplitude, divided by $(-\beta_2)^n$, in the minimal subtraction scheme.

Callan-Symanzik equation \fiftyfive\ and \bone\ imply the following
equation for $Q_n$
$$\eqalign{0&=(n-1)[Q_n-Q_{n-1}]-\delta {\partial
Q_n\over\partial \delta}\cr&+{\beta_3\over\beta_2^2}(n-
2)\delta Q_{n-2}-{\beta_4\over\beta_2^3}(n-3)Q_{n-3}+...
\cr&-\gamma_2\delta Q_{n-2}+\gamma_3\delta^2 Q_{n-3}-...
}\eqn\bthree$$
Now $Q_n$ is a slowly varying function of $n$. Then, in
leading order of $n$ and fixed $x=\delta n$ one obtains the
following differential equaton for $Q_n$
$$0=n{\partial Q_n\over\partial n}-\delta{\partial Q_n
\over \partial \delta}+{\beta_3\over\beta_2^2}xQ_n.
\eqn\bfour$$
Solving \bfour\ one obtains the solution of \bthree\ as
$$Q_n=F(x)\exp\left\{{\beta_3\over\beta_2^2}x\log n\right\}[
1+O(1/n)],
\eqn\bfive$$
where $F(x)$ is to be determined by the inital conditions.

Note that all the momentum dependence is concentrated in
$F(x)$. Nevertheless, since $F(x)$ is dimensionless, it
becomes independent of the momentum components after the
substitution $p_0=|\bp|^\alpha$. The zero of $F(x)$, as a
function of $x$ determines the value of $\alpha_n$. As
$x=D\rho/3$, it provides solutions at fixed $\rho$, as
required by the existence of a nontrivial asymptotic value
for the critical exponent.

The initial conditions fix the form of $F(x)$. In fact
iterating the expanded form \bthree, with
$\beta_3=\beta_4=...=\gamma_2=\gamma_3=...=0$ we obtain
$$F(x)=\sum_{m=0}^{n-1} {x^m\over m!}q^m_{m+1}.\eqn\bsix$$
As we mentioned before, the coefficient $q^m_{m+1}$ is the finite
part of the $m$ loop term, divided by $(-\beta_2)^m$, in the minimal
subtraction scheme.

 It is well known that in a $\phi^4$ field
theory the asymptotic $m$ dependence of $m$ loop
contributions to  Green's functions $\Gamma_n\sim(-1)^m
m!a^{-m}m^kb$.\refmark\brezin\ Consequently, expansion \bsix\ of
$F(x)$ has a finite radius of convergence, $a$. The improved convergence of
series \bsix\ compared to that of the $\epsilon$-expansion is due to the
extracted multipliers $n^m$, contained in the power $x^m$.

The finite radius of convergence, and the fact that the
singularities of the expansion series \bsix\ are at $x<0$
(the series is alternating) allows a conformal transformation
of the variable $x$ such that all values $0<\rho<2$ lie in
the domain of convergence.

Unfortunately, due to the complicated form of the propagator of the effective
theory we have not been able to determine parameter $a$, controlling the radius
of convergence yet. We know, however that $F(x)$ must have a zero between
$0<\rho<2$.
\section{Critical exponents in first approximation}

The leading order term of integral equation \extrathree\ vanishes if
$\alpha$ is the root of the subtracted $n$-loop contribution.
As we have seen it earlier, at large $n$ this is equivalent to
finding the root of the $n$ loop contribution as a function
of $x=n(3\alpha-D)\equiv n\delta$. Then $\nu=1/(2-3x/D)$.
Since the $n$-loop contribution is proportional to $F(x)$, we
have to find a root of $F(x)$. Then a series of critical
exponents, $\nu_k$, alternative to the series obtained in
Sec. 3, can be obtained by approximating $F(x)$ by its terms
up to $x^k$.

In particular, for $k=1$ we obtain $x_1=-1/q^1_2$, where
$q^1_2$ is the ratio of the minimally subtracted finite term
and of the pole term of the subtracted two loop self-energy
diagram. Let us determine $q^1_2$. It is simpler to use old-
fashioned perturbation theory (the amplitude integrated over
timelike momentum components). The subtracted two loop
amplitude, up to irrelevant multipliers, is
$$\eqalign{\Gamma^{(2)}&=\int {d^Dp_1\over
\bp_1^\alpha}{d^Dp_2\over\bp_2^\alpha}
\bigg\{{1\over[\bp^\alpha_1+\bp^\alpha_2+(\hat{\bf1}-\bp_1-
\bp_2)^\alpha]^2-1}\cr&-{1\over[\bp_1^\alpha+\bp_2^\alpha
+(\bp_1+\bp_2)^\alpha]^2}\bigg\}.}\eqn\bten$$
In \bten, as required, we substituted $q_0^2=-\bq^{2\alpha}$
for the components of the external momentum.  Furthermore, we
factored $\bq^{2D-4\alpha}$ out. $\hat{\bf1}$ is a unit vector.

It is comparatively easy to separate the pole term from
\bten, in variable $\delta$. We obtain $$\Gamma^{(2)}_{\rm
pole}= {\Omega_D\over\delta}
\int_0^1d\xi\int d\Omega_D{\xi^{2D/3-1}\over [1+\xi^{D/3}
+(1+\xi^2+2\xi\cos\theta)^{D/6}]^4} .\eqn\beleven$$
The finite part of \bten, taken at $\alpha=D/3$ is
$$\eqalign{\Gamma^{(2)}_{\rm finite}&=\int {d^Dp_1\over
\bp_1^{D/3}}{d^Dp_2\over\bp_2^{D/3}}
\bigg\{{1\over[\bp^{D/3}_1+\bp^{D/3}_2+(\hat{\bf1}-\bp_1-
\bp_2)^{D/3}]^2-1}\cr&-{1\over[\bp_1^{D/3}+\bp_2^{D/3}
+(\bp_1+\bp_2)^{D/3}]^2}-{\theta(|\bp_1|-1)\theta(
|\bp_2|-1)\over[\bp_1^{D/3}+\bp_2^{D/3}
+(\bp_1+\bp_2)^{D/3}]^4}\bigg\}\cr&+{\Omega_D\over3}
\int_0^1d\xi\int d\Omega_D\xi^{2D/3-1}\bigg\{
{5x^{2D/3-1}\ln x\over [1+\xi^{D/3}
+(1+\xi^2+2\xi\cos\theta)^{D/6}]^4}
\cr&-{4x^{D/3}\ln
x+2\ln(1+x^2+2x\cos\theta)(1+x^2+2x\cos\theta)^{D/6}\over
[1+\xi^{D/3}
+(1+\xi^2+2\xi\cos\theta)^{D/6}]^5}.}\eqn\btwelve$$

We have evaluated the ratio $x=-\Gamma_{\rm pole}^{(2)}
/\Gamma^{(2)}_{\rm finite}$ for $D=2$. We obtained
$x=.402 $   , resulting in the prediction
for the critical exponent,
$\nu=.717$, very similar to the one obtained by our first method. Furthermore
we examined the limit
$D\rightarrow3$ as well. The first term of $\Gamma_{\rm finite}^{(2)}$ develops
a pole at $D=3$. Using the standard notation $\epsilon=3-D$,
one obtains $\Gamma_{\rm pole}^{(2)}
/\Gamma^{(2)}_{\rm finite}=-\epsilon/3$, and $\nu=1/(2-\epsilon/3)$, in
agreement with the leading order result from $\epsilon$-
expansion\Ref\wilson{K.G. Wilson and J. Kogut \journal Phys.
Rep. & C12 (74) 75.}.

\chapter{ Summary}
A modified perturbation expansion was introduced in this paper. The expansion
has the virtue of providing a potentially convergent series of approximants for
the critical exponent $\nu$. The subsequent approximations are determined by
subsequent orders of a modified perturbation expansion, in which the infrared
behavior of the propagator is softened. The condition of relativistic
invariance fixes the propagator uniquely.

The results of the first part of this paper are identical to those of a
previous work of ours,\refmark\suranyi\ in which we arrived at the approximants
to the critical exponent taking a different route.  In that paper we obtained
a slightly different form of  \extratwo,
$$E^2(\bq)=\bq^2+m^2+\Sigma(q_0,\bq)
\bigg|_{q_0^2=
E^2(\bq)},\eqn\extranew$$
which was interpreted as an integral equation for $E^2(\bq)$. Then a solution
of nonrelativistic infrared behavior, $E^2(\bq)\sim c\bq^\alpha$, was obtained,
where $\alpha$ coincides with the value of $\alpha$ obtained in the current
paper. In Ref.[\suranyi] it was assumed that the self-consistency of \extranew\
requires the vanishing of coefficient $c$ in infinite order.

The results of the second half of our paper are even more interesting. The
modified perturbation theory defines an effective field theory in the infrared
domain. Large orders of  that effective field theory can be investigated
through the renormalization group. The k-loop approximation to the solution of
the renormalization group equation was shown to provide a series of {\bf finite
radius of convergence}, the zero of which also provides an approximation to the
correlation exponent.

The
prescription, we provided for the determination of the
correlation exponent in $n$ loop order uses the
$n$ loop terms alone. This is the very circumstance which
makes
the universality of the critical exponent (its independence
from
the coupling constant) possible. In the
$\epsilon$-expansion approach the dependence on the coupling constant
is
eliminated by the limit $g\rightarrow g_c(\epsilon)$, where $g_c(\epsilon)$ is
the fixed point of
the
renormalization group transformation. Here we approach the
critical surface at an arbitrary value of the coupling
constant,
by adjusting the bare mass. Thus, $\nu$ can be independent of $g$ only if in
every order it is determined by the highest loop
contribution
alone.

Finally, we discuss the extension of the
results of this paper to the $O(N)$ model. The only difference a
calculation in
the $O(N)$ model would make is that the contribution of
individual
diagrams would be multiplied by an $N$ dependent multiplier.
There is only one diagram contributing to the equation
determining $\alpha_n$ in two and
three loop order (Note that we have chosen the subtraction
scheme in which momentum independent insertions on propagators cancel). Thus
only in the four or higher loop orders
(where there are four or more topologically different
diagrams)
would we get a $N$-dependent $\nu$. This, is certainly not a
fatal flow, since the dependence of $\nu$ on $N$ is weak. As far as the $N$
dependence of the critical exponent is concerned the situation is similar in
the renormalization group approach. This is a little more surprising, because
for $N=1$ (but not for other $N$) at $D\simeq3$ the result of the
$\epsilon$-expansion is reproduced in the two loop approximation.

The author is indebted to L.C.R. Wijewardhana, for discussions, to Beth
Basista, for the critical reading of the manuscript, and to Michael Ma, for
calling his attention to Ref.[\fishertwo]. This work has been supported in part
by the US Department of Energy, through Grant DE-FG02-84ER40153. The support
of
the Ohio Supercomputer Center for computing time and of the Research Council of
the University of Cincinnati for computer hardware is gratefully acknowledged.

\refout
\bye